\documentclass[letterpaper, 10 pt, conference]{ieeeconf} 

\IEEEoverridecommandlockouts  
\usepackage{cite}
\usepackage{amsmath,amssymb,amsfonts}
\usepackage{algorithmic}
\usepackage{graphicx}
\usepackage{caption}
\usepackage{bm}
\usepackage{float}
\usepackage{hyperref}
\usepackage{mathtools} 
\usepackage{textcomp}
\usepackage{xcolor}
\usepackage{multirow}

\def\BibTeX{{\rm B\kern-.05em{\sc i\kern-.025em b}\kern-.08em
    T\kern-.1667em\lower.7ex\hbox{E}\kern-.125emX}}

\begin{document}

\title{Global Sensitive-Based Input Shaping for UAV-Payload Precision Motion Control}

\author{Karan Baker$^{1}$, Sanjay Maharjan$^{1}$, Tariq Hlayel$^{1}$, Oladapo Ogunbodede$^{2}$, Dutch Dunphy$^{1}$, and Adrian Stein$^{1}$
\thanks{$^{1}$K. Baker, S. Maharjan, T. Hlayel, D. Dunphy, and A. Stein are with the Department of Mechanical and Industrial Engineering, Louisiana State University, LA 70803, USA. {\tt\small(email: \{kbake54,smaha12,thlaye1,ddunph3,astein\}@lsu.edu)}.}
\thanks{$^{2}$ O. Ogunbodede is with GE Aerospace, Cincinnati OH, USA. 
{\tt\small(email: dapo.ogunbodede@gmail.com)}
This work was supported by LaSPACE under 80NSSC20M0110 and 80NSSC25M7120.}}
\maketitle

\begin{abstract}
This work presents a comprehensive analysis and design of global sensitivity-based input shapers for a 3D Unmanned Aerial Vehicle-payload system, emphasizing robustness against uncertainties in payload mass and rope length. The proposed approach also leverages the Shapley value concept in controller design to systematically account for uncertainties, thereby reducing the controller’s sensitivity to unknown parameters. To validate the effectiveness of the methodology, numerical simulations are conducted, comparing the proposed controller against non-robust, robust, and minimax designs. The results demonstrate that the standard global sensitivity or Shapley-based input shapers improve performance and offer a promising framework for uncertainty-aware control in aerial payload transport.
\end{abstract}
\section{INTRODUCTION}
The rapid expansion of aerial robotics has fueled interest in unmanned aerial vehicles (UAVs) transporting suspended payloads. Such systems combine high maneuverability with mechanical simplicity, allowing them to perform tasks in environments that are inaccessible or impractical for ground vehicles \cite{li_autotrans_2023, shelare_payload_2024, mahesh_fire_2023}. Typical applications include last-mile delivery \cite{akhtar_path_2022, qian_guidance_2020}, cooperative transport of heavy loads using multiple UAVs \cite{han_controller_2022, james_cooperative_2024}, and precision construction or additive manufacturing tasks where aerial robots act as mobile manipulators \cite{hunt_3d_2014, nettekoven_3d_2021}. These diverse cases highlight the versatility and promise of UAV-payload systems in advancing modern logistics, infrastructure, and emergency response.  

Despite these benefits, UAV-payload transport remains fundamentally difficult to control due to its underactuated and nonlinear nature. The suspended load introduces additional degrees of freedom, leading to complex coupled dynamics between the vehicle and the payload \cite{yang_energy-based_2020, zhu_modeling_2025}. Even small deviations in the UAV’s trajectory can excite oscillatory payload motion, which feeds back into the vehicle dynamics. This bidirectional coupling often results in residual swing that undermines trajectory tracking accuracy, reduces payload safety, and, in extreme cases, destabilizes the entire system \cite{yazdannik_novel_2024, geronel_adaptive_2025}. These challenges are further compounded by parameter uncertainties, such as variations in the payload mass or rope length, which are often unknown prior to flight or subject to change during operation \cite{imran_adaptive_2024}.  

To suppress payload swing while preserving tracking performance, many control methodologies have been proposed. Passivity-based controllers exploit the energy structure of the system to ensure stability under specific operating conditions \cite{mohammadi_passivity-based_2022}. Nonlinear feedback strategies have been shown to attenuate oscillations while remaining relatively simple to implement in real time \cite{ramli_pid_2025, stein_aruco_2024}. Adaptive controllers and sliding mode designs provide robustness by adjusting to uncertainties or enforcing invariance despite modeling errors \cite{eltayeb_integral_2022, muthusamy_real-time_2022}. While each of these approaches demonstrates clear advantages, they are typically tuned for nominal operating points or bounded parameter variations. As a result, their performance may degrade significantly when the system is subject to broad or unstructured uncertainties.  

Input shaping offers a particularly attractive solution for payload oscillations in uncertain systems. By modifying the reference command through convolution with a sequence of impulses, input shaping filters reduce residual vibrations without requiring additional feedback or high controller bandwidth. Early formulations, such as zero-vibration, zero-vibration-derivative, or minimax shapers, were later extended to robust designs that explicitly account for parameter variations \cite{singh_closed-form_2007, singh_optimal_2009}. More recently, input shaper (IS) formulations have been introduced to balance vibration suppression and robustness while retaining closed-form solutions \cite{stein_minimum_2023}. Parallel to these developments, global sensitivity analysis (GSA) methods have emerged as powerful tools for quantifying how system uncertainties affect performance, providing a systematic framework for designing controllers that are robust \cite{stein_global_2022,stein_shapley_2025}. 

This work demonstrates the feasibility of applying a GSA-based input shaper to a UAV-payload system. Furthermore, a GSA input shaper incorporating Shapley value constraints is proposed to analyze variations in the payload's swinging angles. Originally introduced in cooperative game theory, the Shapley value offers a principled method for fairly attributing the contribution of each input variable to the overall system response under uncertainty~\cite{kuhn_17_1953}. Two key parameters, the payload’s length and mass, are treated as uncertain in this work.

The remainder of this manuscript is organized as follows. Section~\ref{sec:methodology} details the methodology for the UAV-payload modeling and implementation of GSA-based input shapers on a $3$D UAV-payload system. Section~\ref{sec:results} presents simulation results comparing different controller designs, including non-robust, robust, minimax, and the proposed GSA-based approach. Section~\ref{sec:Experiment} outlines future experimental validation of the GSA-based input shaper in comparison to a shaper-less "pulse" profile and classic input shaping pipelines. Finally, Section~\ref{sec:conclusions} provides a concise summary of the main findings and contributions.
%
%
%
\section{METHODOLOGY}
\label{sec:methodology}
This section provides an overview of the UAV-payload dynamics, input shaper design, and input shaper optimization via a standard GSA formulation or the Shapley value theorem.
%
%
%
\subsection{UAV-Payload Model}
\begin{figure}
    \centering
    \includegraphics[width=0.3\textwidth]{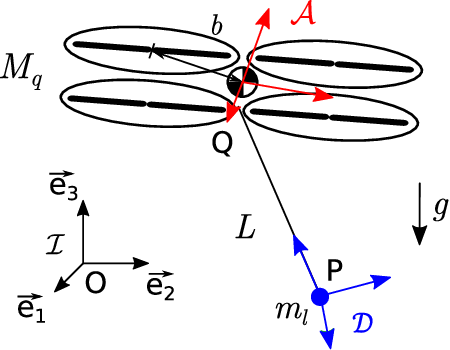} 
    \caption{Inertial and body frames of UAV-payload system.}
    \label{fig:01_UAV_with_Payload}
\end{figure}

\begin{figure}
    \centering
    \includegraphics[width=0.4\textwidth]{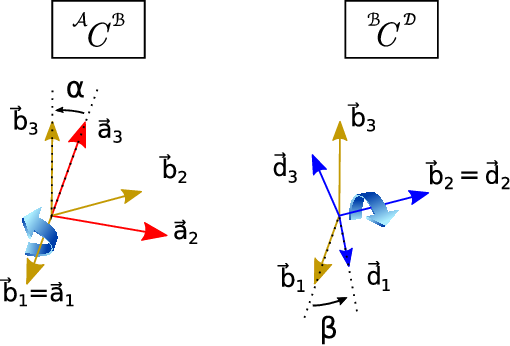} 
    \caption{Transformation of the reference frames $^\mathcal{A}C^\mathcal{B}$ and $^\mathcal{B}C^\mathcal{D}$.}
    \label{fig:02_Transformations}
\end{figure}

The UAV is modeled as a point mass with thrust forces generated by four rotors, each located a distance $b$ from the UAV’s center of mass. A point mass payload $m_l$ is attached via a rigid cable of length $L$, such that the UAV's position is $Q=[x_q,y_q,z_q]^T$ and the payload's position is $P=[x_l,y_l,z_l]^T$. Fig.~\ref{fig:01_UAV_with_Payload} illustrates the UAV-payload system with the inertial frame $\mathcal{I}$ and the body frames of the UAV and payload defined as $\mathcal{A}$ and $\mathcal{D}$, respectively. The Euler angles denote the orientation of the body frame $\mathcal{A}$ to the inertial frame $\mathcal{I}$ and are given by the angles $\phi$, $\theta$, and $\psi$, which represent the roll, pitch, and yaw of the UAV. 
The gravitational acceleration is set to $9.81$ m/s\textsuperscript{2} and \mbox{$b=0.21$ m}. For simplicity, $c(\cdot)$ and $s(\cdot)$ are used for $\cos(\cdot)$ and $\sin(\cdot)$. The transformation matrices for the UAV are:
\begin{align}
    &\mathbf{R}_x(\phi) = 
    \begin{bmatrix}
    1 & 0 & 0 \\
    0 & c\phi & -s\phi \\
    0 & s\phi & c\phi
    \end{bmatrix};\
    \mathbf{R}_y(\theta) = 
    \begin{bmatrix}
    c\theta & 0 & s\theta \nonumber\\
    0 & 1 & 0 \\
    -s\theta & 0 & c\theta
    \end{bmatrix},\\
    &\mathbf{R}_z(\psi) = 
    \begin{bmatrix}
    c\psi & -s\psi & 0 \\
    s\psi & c\psi & 0 \\
    0 & 0 & 1
    \end{bmatrix},
\end{align}
which results in the transformation from the UAV body to inertial frame as:
\begin{align}
    ^\mathcal{I}C^\mathcal{A} = \mathbf{R}_{z}(\psi)\mathbf{R}_{y}(\theta)\mathbf{R}_{x}(\phi).
\end{align}
The transformation matrices from the UAV body frame $\mathcal{A}$ to the payload's reference frame $\mathcal{D}$ are:
\begin{subequations}
    \begin{align}
        \begin{bmatrix}
        \vec{a}_1\\
        \vec{a}_2\\
        \vec{a}_3\\
        \end{bmatrix}
        & =
        \begin{bmatrix}
        1 & 0 & 0\\
        0 & c\alpha & -s\alpha\\
        0 & s\alpha & c\alpha
        \end{bmatrix}
        \begin{bmatrix}
        \vec{b}_1\\
        \vec{b}_2\\
        \vec{b}_3\\
        \end{bmatrix}, \label{eq:transformation_matrix_ACB}\\
        \begin{bmatrix}
        \vec{b}_1\\
        \vec{b}_2\\
        \vec{b}_3\\
        \end{bmatrix}
        & =
        \begin{bmatrix}
        c\beta & 0 & s\beta\\
        0 & 1 & 0\\
        -s\beta & 0 & c\beta
        \end{bmatrix}
        \begin{bmatrix}
        \vec{d}_1\\
        \vec{d}_2\\
        \vec{d}_3\\
        \end{bmatrix}, \label{eq:transformation_matrix_BCD}
    \end{align}
\end{subequations}
where $\alpha$ and $\beta$ are the swinging angles of the payload. The payload's twist is not modeled in this work. Note that $^{\mathcal{A}}C^{\mathcal{D}} = {}^{\mathcal{A}}C^{\mathcal{B}} \; {}^{\mathcal{B}}C^{\mathcal{D}}$. Furthermore, these transformations are illustrated in Fig.~\ref{fig:02_Transformations}.
The payload's position and velocity in the inertial reference frame can then be written as:
\begin{subequations}  
    \begin{align}
        ^\mathcal{I}\mathbf{r}_{P,O} = ^\mathcal{I}\mathbf{r}_{Q,O} + ^\mathcal{I}C^\mathcal{A} \: ^\mathcal{A}C^\mathcal{D}
        \begin{bmatrix}
            0\\
            0\\
            -L
        \end{bmatrix},\\
        \rightarrow ^\mathcal{I}\mathbf{v}_{P,O} = ^\mathcal{I}\frac{d}{dt}\left(^\mathcal{I}\mathbf{r}_{P,O}\right),
    \end{align}
\end{subequations}
To calculate the rotational kinetic energy of the UAV, the derivative of the Euler angles to the body angular velocities is given as~\cite{wang_dynamics_2016}:
\begin{align}
   \left[^\mathcal{I}\bm{\omega}^\mathcal{A} \right]_\mathcal{A} = \begin{bmatrix}
        p\\
        q\\
        r 
    \end{bmatrix}
    =
    \begin{bmatrix}
        1 & 0 & -s\theta\\
        0 & c\phi & c\theta s\phi\\
        0 & -s\phi & c\theta c\phi
    \end{bmatrix}
    \begin{bmatrix}
        \dot{\phi}\\
        \dot{\theta}\\
        \dot{\psi}
    \end{bmatrix}.
\end{align}
The Lagrangian $\mathcal{L} = T_{sys} - U_{sys}$ yields the equations of motion through:
\begin{align}
    \frac{d}{dt} \left( \frac{\partial \mathcal{L}}{\partial \dot{\gamma}_i} \right) - \frac{\partial \mathcal{L}}{\partial \gamma_i} = \mathbf{\Gamma}_i,
\end{align}
with $\gamma$ being the generalized coordinate and \mbox{$\mathbf{\Gamma}_i = [F_x, F_y, F_z, \tau_x, \tau_y, \tau_z]^T$} representing generalized forces, which are the external thrust and moments generated by the rotors. The kinetic energy $T_{sys}$ of the coupled system includes the translational motion of the UAV, the payload swing, and the rotational energy of the UAV~\cite{ogunbodede_load_2022}:
\begin{align}
    T_{sys} = \frac{M_q}{2}||^\mathcal{I}\mathbf{v}_{Q,O}||^2 + \frac{m_l}{2}||^\mathcal{I}\mathbf{v}_{P,O}||^2 \nonumber\\  
    + \frac{1}{2} \left[^\mathcal{I}\bm{\omega}^\mathcal{A} \right]_\mathcal{A}^T \left[\mathbf{I}_Q\right]_\mathcal{A} \left[^\mathcal{I}\bm{\omega}^\mathcal{A} \right]_\mathcal{A},
\end{align}
where $\left[\mathbf{I}_Q\right]_\mathcal{A}$ is a $3\times3$ diagonal matrix consisting of the moments of inertia $I_{xx}$, $I_{yy}$, and $I_{zz}$. In this work \mbox{$I_{xx}=0.0465$}, $I_{yy}=0.0451$, and $I_{zz}=0.0697$. The mass of the UAV is $M_q=3$ kg. The potential energy of the system is:
\begin{align}
    U_{sys} = \left(M_q + m_l\right)gz_q - m_l g L c(\alpha)c(\beta).
\end{align}
In this work the payload's mass $m_l$ and cable length $L$ are assumed uncertain with a uniform distribution of \mbox{$m_l \sim \mathcal{U}\left[0.2,1\right]$ kg} and $L \sim \mathcal{U}\left[0.2,1.8\right]$ m. Uniform distributions are used in these simulations because parameter bounds are set as given information, but this approach can utilize other probability distributions. The controller design of the UAV is based on work presented by Ogunbodede and Singh~\cite{ogunbodede_load_2022}.
%
%
%
\subsection{Input Shaper Design}
\label{sec:input_shaper_design}
Input shaping is a feedforward control technique that modifies command signals by convolving them with designed impulses to reduce residual vibrations in flexible systems. This work focuses on different input shaping designs to reduce the swinging angles $\alpha$ and $\beta$ of the payload during a point-to-point maneuver. Singh~\cite{singh_optimal_2009} defined an input shaper with one delay as:
\begin{subequations}
\begin{align}
     G_{\mathrm{nonrob}}(s) = A_0 + A_1 e^{-s T}, \label{eq:G_nonrob}\\
    A_0 = \frac{e^{\left(\frac{\zeta\pi}{\sqrt{1-\zeta^2}}\right)}}{1+e^{\left(\frac{\zeta\pi}{\sqrt{1-\zeta^2}}\right)}};\:\: A_1 = 1-A_0;\:\: T = \frac{\pi}{\omega_d},
\end{align}
\end{subequations}
where $A_0$ and $A_1$ are the magnitudes of the proportional signal and $T$ is the delay time. $\omega_d$ is the damped natural frequency defined as $\omega_d=\omega_n\sqrt{1-\zeta^2}$. This design places one set of zeros at the nominal location of the underdamped system poles and is referred to as a non-robust input shaper in this work. The nominal location is defined by modeling the system with the expected values of the uncertain parameters, chosen as payload mass \mbox{$m_{l,nom}=0.6$ kg} and cable length $L_{nom}=1$ m.

\begin{figure}
\centering
	\includegraphics[width=0.48\textwidth]{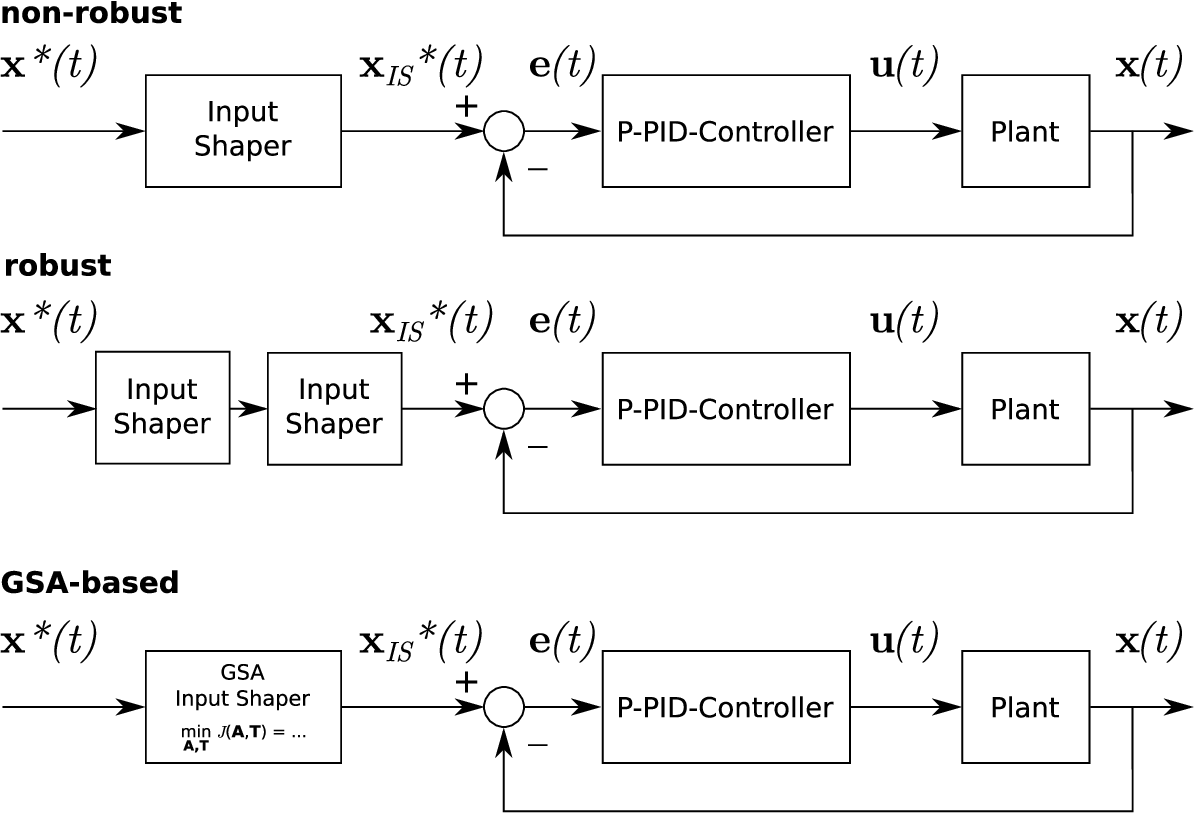}
	\caption{Block diagram of different input shaper designs.}
 \label{fig:03_Block_Diagram}
\end{figure}

Another shaper design is the robust input shaper which is selected to place two sets of zeros at the nominal location of the underdamped poles of the system resulting in:
\begin{align}
    G_{\mathrm{rob}}(s) = \left(A_0 + A_1 e^{-s T}\right)^2. \label{eq:G_rob}
\end{align}
A third option is a minimax input shaper which has the following structure~\cite{singh_optimal_2009}:
\begin{subequations}
    \begin{align}
        G_{\mathrm{minimax}}(s) = A_0 + A_1 e^{-s T_1} + A_2 e^{-s T_2},\\
        A_0 = \frac{2\left(1+\cos\left(\frac{\omega_L\pi}{\omega_d}\right)\right)}{5+4\cos\left(\frac{\omega_L\pi}{\omega_d}\right)-\cos\left(\frac{2\omega_L\pi}{\omega_d}\right)}\nonumber,\\
        A_1 = 1-2A_0;\:\: A_2 = A_0;\:\: T_1 = \frac{\pi}{\omega_d};\:\:T_2 = 2T_1,
    \end{align}
\end{subequations}
where $\omega_d=(\omega_U+\omega_L)/2$ with $\omega_L$ and $\omega_U $ being user-defined. In this case $\omega_L = 0.7\omega_d$ and $\omega_U = 1.3\omega_d$ are chosen for each identified $\omega_d$ for $\alpha$ and $\beta$, respectively. The robust and minimax input shapers, while seemingly similar, are quite distinct in effect. The robust shaper, by adding a second set of the same zero, can widen the notch at the location of the zero, which reduces the residual energy if the poles shift away from the zero's location. The minimax shaper provides more flexibility to splitting the original command signal, which improves minimization of residual energy away from the nominal poles. \\
\indent In this work the desired position and velocity are shaped. Fig.~\ref{fig:03_Block_Diagram} illustrates the different input shaper options such as non-robust, robust, and GSA-based IS. $\mathbf{x}^*(t)$ and $\mathbf{x}^*_{IS}(t)$ are the unshaped and shaped reference inputs while $\mathbf{x}(t)$ are the states. $\mathbf{e}(t)$ are the errors between the desired and actual states, and $\mathbf{u}(t)$ are the inputs for each state.
Since $\alpha$ and $\beta$ are dynamically coupled, both shapers are connected in sequence. Therefore a non-robust, robust, and minimax IS have $4$, $9$, and $9$ switches, respectively.
%
%
%
\subsection{Shapley Value Solution Concept}
For a model with $n$ uncertain variables, Shapley values can be used as a solution to a coalition game, defined in~\cite{kuhn_17_1953} as:
\begin{align}
    Sh_{\kappa} &= \frac{1}{n}\sum_{{h\subseteq -\{\kappa\}}} \binom{n-1}{|h|}^{-1} \left(w(h\cup \{\kappa\}) - w(h)\right), \label{eq:Shapley_version}
\end{align}
where $w(\cdot)$ represents the worth of a coalition, and $-\{\kappa\}$ denotes the set of indices $\{1,\dots,n\}$ excluding $\kappa$. Here, $w(h \cup \{\kappa\})$ corresponds to a coalition containing variable $\kappa$, whereas $w(h)$ excludes it. Shapley values measure the marginal contribution of variable $\kappa$ within the coalition. For probabilistic systems, the coalition value is expressed as~\cite{iooss_shapley_2019}:
\begin{align}
    w(h)=\frac{Var\left(\mathbb{E}[Z|\mathbf{X}_h]\right)}{Var(f)},\label{eq:Shapley_c_u}
\end{align}
where $\mathbf{X}=(\mathbf{X}_1,\dots,\mathbf{X}_d)$ is a set of independent continuous inputs, and $\mathbf{X}_h$ denotes the subset of variables indexed by $h\subseteq \{1,\dots,n\}$. $f$ is the scalar model output. For multiple outputs, a superscript $o$ can be added to indicate the output of interest, i.e., $Sh_{\kappa}^o$. Shapley values have the following characteristics when solving analytically~\cite{iooss_shapley_2019}: each must be non-negative and, for a given model, the sum of all Shapley values must equal $1$.

For the purposes of the problem addressed in this work, the players for this "game" are the system parameters, which are the mass of the payload and the length of the rope. Furthermore, since minimizing the range of the swinging angles through the IS is the goal, the shaper design parameters (the time delays and magnitudes of the delayed components) also have weight as players. 
%
%
%
\subsection{Global Sensitivity-based IS Design}
The aim is to design an input shaper that reduces the range of swinging angles $\alpha$ and $\beta$, defined by the difference of maximum and minimum swinging angle, over the whole maneuver time for uncertain payload mass and cable length. The optimization problem can be formulated as follows:
\begin{subequations}  
    \begin{align}
        \min_{\mathbf{A},\mathbf{T}} \quad &
        J(\mathbf{A},\mathbf{T})
        = \mathbb{E}\left[ R_\alpha(m_\ell,L;\mathbf{A},\mathbf{T})\right]
           \nonumber\\
        & + \mathbb{E}\left[R_\beta(m_\ell,L;\mathbf{A},\mathbf{T})\right] \\
        \text{s.t.}\quad
        & \sum_{i=1}^{9} A_i = 1, \\
        & 0 \leq A_i \leq 1, \quad i=1,\dots,8, \\
        & 0 < T_1 < T_2 < \cdots < T_7  < T_8 = t_f,\\
        & \text{where:} \quad m_l \sim \mathcal{U}\left[0.2,1\right];\:L \sim \mathcal{U}\left[0.2,1.8\right],\\
        & \mathbf{A} = \left[A_1,A_2,\cdots,A_8\right]; \: \mathbf{T} = \left[T_1,T_2,\cdots,T_7\right],
    \end{align}
    \label{eq:optimization_problem}
\end{subequations}
where $\mathbf{A}$ contains all the magnitudes for the IS and $\mathbf{T}$ represent all the switch times. The GSA IS magnitudes $\mathbf{A}$ are enforced to be positive and sum up to $1$ to maintain the unity DC gain. A similar GSA IS design has been shown in work~\cite{stein_global_2022} but there the residual energy was used as a metric. For a fair comparison to established IS design, the final time of the GSA IS matches the end of the robust closed-form IS in~\eqref{eq:G_rob}, here \mbox{$t_f=3.7898$ s}. Furthermore, the range of the swinging angles $\alpha$ and $\beta$ can be written as:
\begin{subequations}  
    \begin{align}
        R_\alpha(m_\ell,L;\mathbf{A},\mathbf{T})
        &= \max_{t \in [0,T_f]}
            \alpha(t; m_\ell,L,\mathbf{A},\mathbf{T}) \nonumber\\
           &- \min_{t \in [0,T_f]}
            \alpha(t; m_\ell,L,\mathbf{A},\mathbf{T}), \\
        R_\beta(m_\ell,L;\mathbf{A},\mathbf{T})
        &= \max_{t \in [0,T_f]}
            \beta(t; m_\ell,L,\mathbf{A},\mathbf{T}) \nonumber\\
           &- \min_{t \in [0,T_f]}
            \beta(t; m_\ell,L,\mathbf{A},\mathbf{T}),
    \end{align}
    \label{eq:range_alpha_and_beta}
\end{subequations}
where $T_f$ is the final maneuver time. This design is referred to as the standard GSA IS. 

Additionally to the standard GSA IS, a Shapley-based GSA IS can be introduced by adding a constraint \mbox{$Sh^{R_\alpha}_L \leq 0.85$} to the optimization problem described in~\eqref{eq:optimization_problem}. For clarification, this constraint could be formulated as the following question: when contending with uncertain parameters $m_l$ and $L$ in the UAV-payload system, what is the influence of $L$ on the range of a swinging angle? To solve this optimization problem, MATLAB's \texttt{fmincon}-function with the interior-point method and $10,000$ Monte Carlo samples is used for the uniformly distributed variables to solve the highly nonlinear problem. As mentioned prior, this approach can be altered to utilize different probability distributions for uncertain parameters.  While the computational cost of the offline design increases proportionally with the number of samples, the input shaper implemented online does not carry on this burden. Each UAV-payload flight is simulated for \mbox{$T_f=15$ s} using MATLAB's ode45 solver, with both absolute and relative tolerances set to $10^{-12}$. The initial conditions for $\mathbf{A}$ and $\mathbf{T}$ are the values from the robust IS. 
%
%
%
\section{RESULTS}
\label{sec:results}
The starting point of the UAV is at $\mathbf{x}_0 = [0,0,0]^T$ and the final desired position is $\mathbf{x}_f = [2,4,10]^T$. Initially, all state variables are set to $0$. The controller gains can be seen in Table~\ref{tab:controller_gains_UAV}.
\begin{table}
\centering
\caption{Controller Gains for UAV}
\label{tab:gains}
\begin{tabular}{lcccc}
\hline
& \textbf{$K_p$} & \textbf{$K_{pv}$} & \textbf{$K_{iv}$} & \textbf{$K_{dv}$} \\
\hline
X-axis & 1.000 & 0.900 & 0.200 & 0.005\\
Y-axis & 1.000 & 0.900 & 0.200 & 0.005\\
Z-axis & 1.000 & 0.400 & 0.070 & 0.000\\
\hline
Roll ($\phi$) & 6.000 & 0.300 & 0.100 & 0.005\\
Pitch ($\theta$) & 6.000 & 0.300 & 0.100 & 0.005\\
Yaw ($\psi$) & 3.000 & 0.240 & 0.100 & 0.000\\
\hline
\end{tabular}
\label{tab:controller_gains_UAV}
\end{table} 
%
%
%
\subsection{System Identification}
\begin{figure}
\centering
	\includegraphics[width=0.4\textwidth]{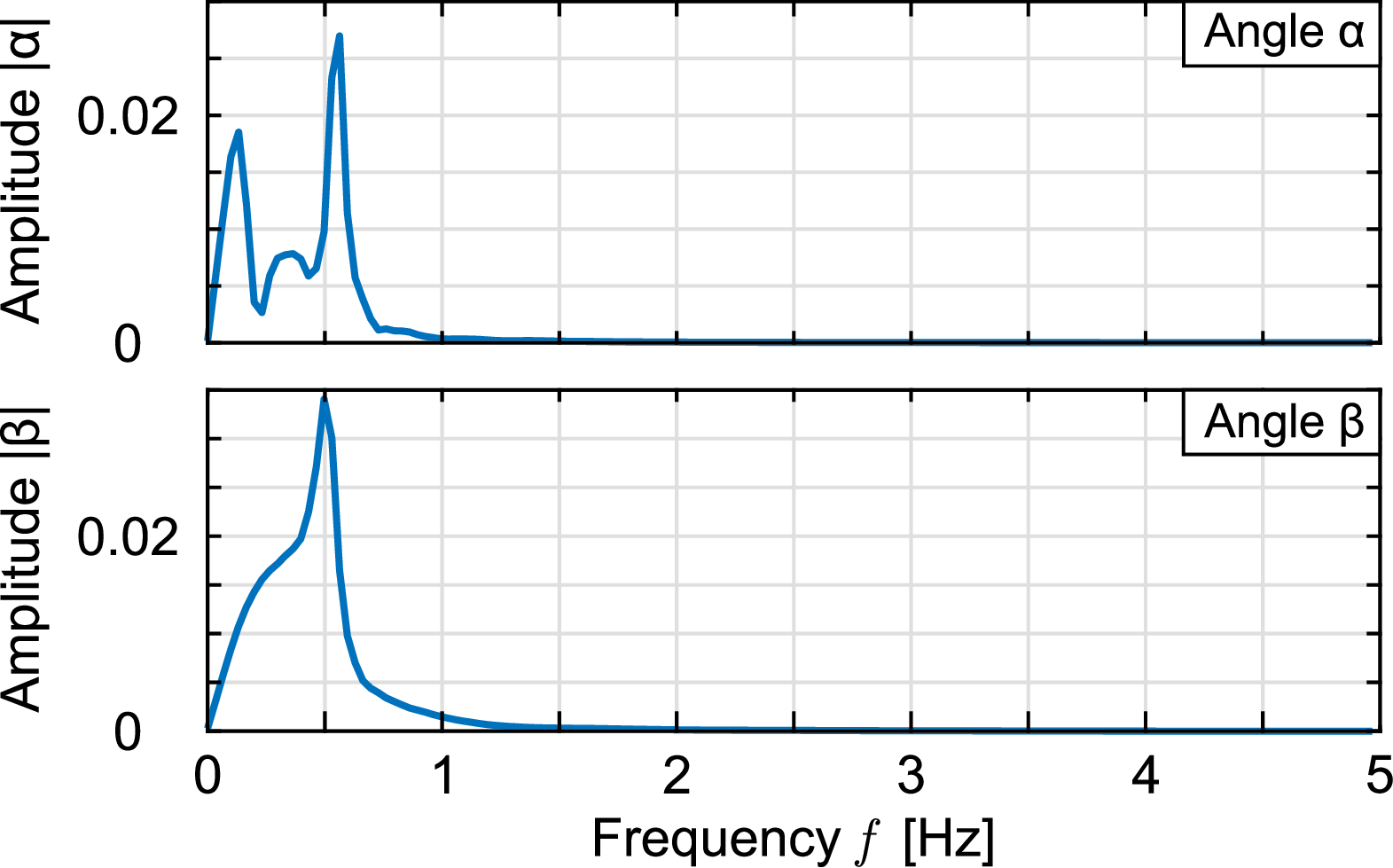}
	\caption{Single-sided amplitude spectrum of swinging angles $\alpha$ and $\beta$.}
 \label{fig:04_FFT_alpha_beta_final}
\end{figure}

To successfully implement and create an input shaper, the system needs to be identified. The parameters which are needed for the IS design are the natural frequency $\omega_n$ and the damping ratio $\zeta$. Fig.~\ref{fig:05_No_IS_final} (B) illustrates the time signal of $\alpha$ and $\beta$ for a case where no IS is applied. Fig.~\ref{fig:04_FFT_alpha_beta_final} provides a single-sided amplitude spectrum of these swinging angles $\alpha$ and $\beta$. The natural frequencies for $\alpha$ and $\beta$ are found to be $0.563$ and $0.497$ Hz. The damping ratios are $0.0330$ and $0.0711$, which are identified via the logarithmic decrement.
%
%
%
\subsection{Classic Input Shaper Comparison}
\begin{figure}
\centering
	\includegraphics[width=0.48\textwidth]{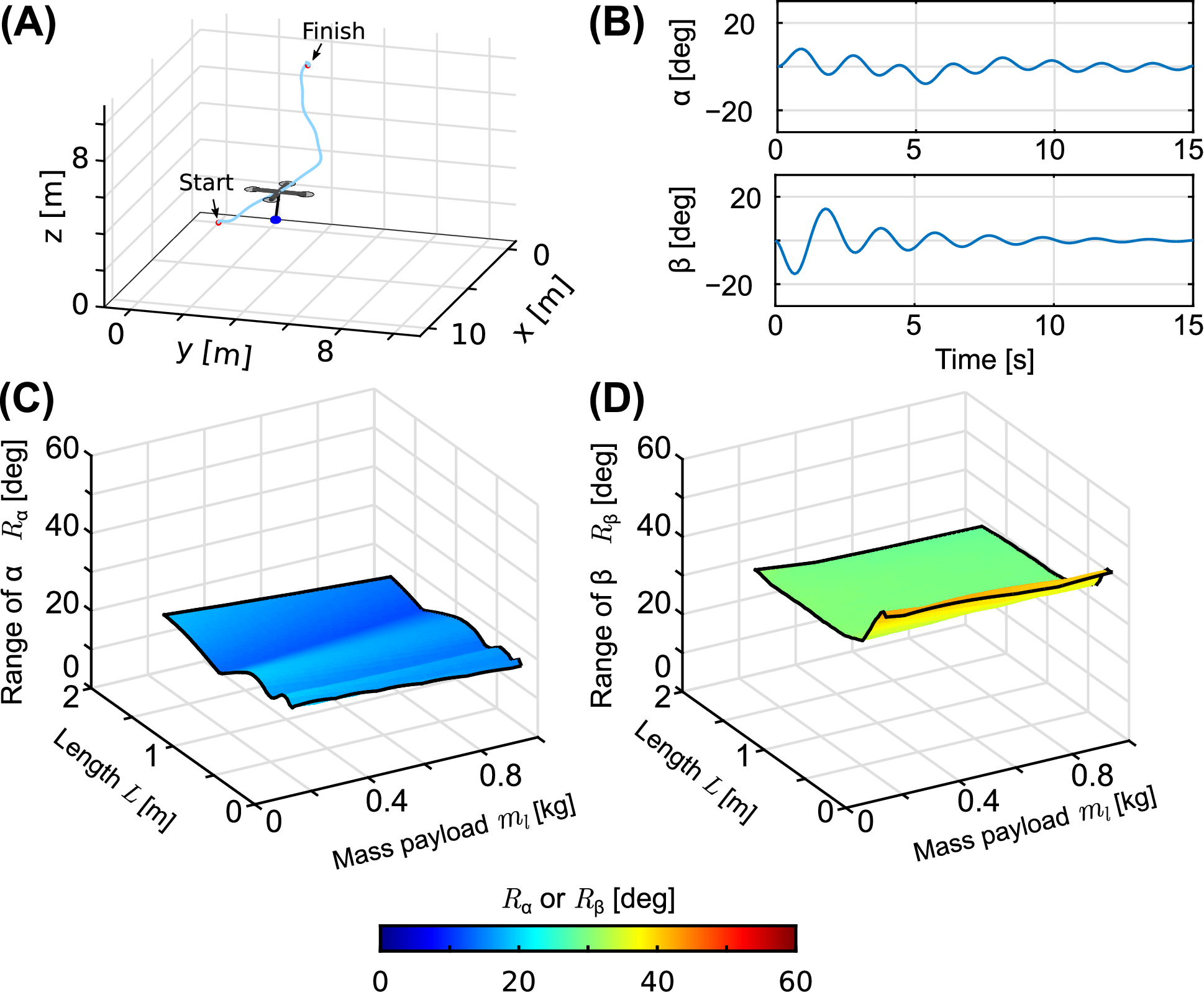}
	\caption{Simulation 
    for pure P-PID control. (A) Flight path in Cartesian coordinates, (B) Swinging angles $\alpha$ and $\beta$. (C)-(D) Range of $\alpha$ and $\beta$.}
 \label{fig:05_No_IS_final}
\end{figure}

This section showcases the performances of the non-robust, robust, and minimax IS. As shown in Fig.~\ref{fig:05_No_IS_final}, the first simulation does not apply an IS to the system. The flight path of the UAV is shown in Fig.~\ref{fig:05_No_IS_final} (A). The results in (B) reveal large oscillations in the system's response. Fig.~\ref{fig:05_No_IS_final} (C) and (D) show the results of multiple maneuvers resulting from a parameter sweep for different $m_l$ and $L$ without IS while plotting the range of both swinging angles. A key observation is that the maximum swinging angles showed a strong dependency on the length $L$, while the payload's mass $m_l$ has a comparatively smaller influence. Such a detail is supported by the effects of length and mass on the dominant mode of a pendulum-like suspended mass. This significant sensitivity to $L$ serves as a strong motivation to design a GSA-based shaper with the Shapley Value Theorem included for desensitizing the system's behavior with respect to variations in $L$.
Note that the non-robust, robust, and minimax IS are designed for the nominal parameters \mbox{$L_{nom}=1$ m} and $m_{l,nom}=0.6$ kg. The final maneuver times for non-robust, robust, and minimax profiles are $1.8949$, $3.7898$, and \mbox{$3.7898$ s}. Naturally, the non-robust IS converges faster with fewer switches compared to the robust and minimax designs. The magnitudes and switch times are shown in Table~\ref{table:magnitudes_and_switch_times_ISs_Case_Benchmark}. The individual values are more evenly distributed for the minimax IS compared to the robust IS, which is most apparent with the maximum magnitudes for the robust shaper and minimax shapers being $0.2462$ and $0.1959$, respectively. 
\begin{table}
\begin{center}
\caption{Magnitudes and switch times for non-robust, robust, and minimax IS.}\label{table:magnitudes_and_switch_times_ISs_Case_Benchmark}
\begin{tabular}{cccc}
Switch Times & non-robust & robust & minimax \\
\hline
$T_1$ & 0.8882 & 0.8882 & 0.8882\\
$T_2$ & 1.0067 & 1.0067 & 1.0067\\
$T_3$ & 1.8949 & 1.7764 & 1.7764\\
$T_4$ & - & 1.8949 & 1.8949\\
$T_5$ & - & 2.0134 & 2.0134\\
$T_6$ & - & 2.7831 & 2.7831\\
$T_7$ & - & 2.9016 & 2.9016\\
$T_8$ & - & 3.7898 & 3.7898\\
\hline
\hline
Magnitudes & non-robust & robust & minimax \\
\hline
$A_0$ & 0.2923 & 0.0854 & 0.0777\\
$A_1$ & 0.2635 & 0.1540 & 0.1234\\
$A_2$ & 0.2336 & 0.1366 & 0.1234\\
$A_3$ & 0.2106 & 0.0694 & 0.0777\\
$A_4$ & - & 0.2462 & 0.1956\\
$A_5$ & - & 0.0546 & 0.0777\\
$A_6$ & - & 0.1110 & 0.1234\\
$A_7$ & - & 0.0984 & 0.1234\\
$A_8$ & - & 0.0444 & 0.0777\\
\hline
\end{tabular}
\end{center}
\end{table}
Table~\ref{table:mean_and_std_classic_shapers} shows that minimax is slightly outperforming the robust design in terms of mean ($\mu$) and standard deviation ($\sigma$) for the ranges $R_\alpha$ and $R_\beta$. It is clear that the robust and minimax design deliver better performance than the pure PID and non-robust profiles, so they can be used for comparison with the GSA input shapers.
\begin{table}
\begin{center}
\caption{Mean and standard deviation of pure P-PID, non-robust, robust, and minimax IS.}\label{table:mean_and_std_classic_shapers}
\begin{tabular}{cccccc}
Metric & P-PID & non-robust & robust & minimax\\
\hline
$\mu(R_\alpha)$ [deg] & 15.7166 & 9.6368 & 8.9806 & 8.7637 \\
$\sigma(R_\alpha)$ [deg] & 1.6721 & 1.4922 & 1.2422 & 1.2303 \\
$\mu(R_\beta)$ [deg] & 31.4199 & 17.6050 & 13.1290 & 12.5919 \\
$\sigma(R_\beta)$ [deg] & 3.4905 & 5.5046 & 5.9510 & 6.3497 \\
\hline
\end{tabular}
\end{center}
\end{table}
%
%
%
\subsection{GSA Input Shaper Comparison}
\begin{figure}
\centering
	\includegraphics[width=0.4\textwidth]{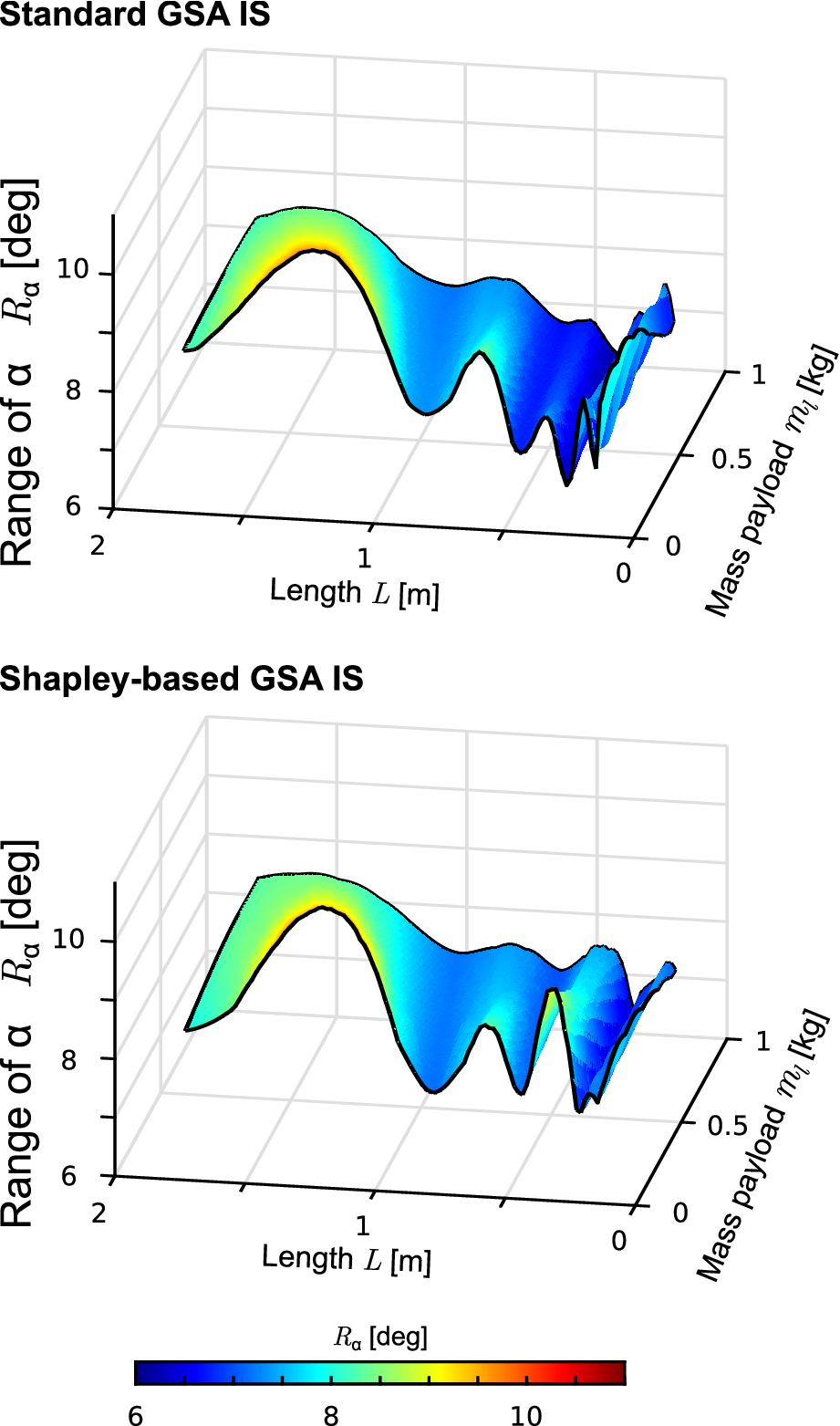}
	\caption{Comparison between standard and Shapley GSA TDFs for range of $\alpha$ for changing payload mass $m_l$ and cable length $L$.}
 \label{fig:06_GSA_final}
\end{figure}

As mentioned previously, the non-robust, robust, and minimax input shapers have $4$, $9$, and $9$ switches, respectively, all with a positive magnitude. For a fair comparison between classic robust/minimax shapers, the standard GSA and Shapley-based GSA IS also have $9$ switches, and the final time of the robust/minimax IS is used for the Shapley-based GSA IS. In Table~\ref{table:magnitudes_and_switch_times_ISs_Case_GSA} the magnitudes and switch times of the standard and Shapley-based GSA IS are compared. While the final times are the same between the GSA shapers, the magnitudes deviate, with more spread among the switches seen for the Shapley-based GSA compared to the standard GSA.
\begin{table}
\begin{center}
\caption{Magnitudes and switch times for standard and Shapley-based GSA IS.}\label{table:magnitudes_and_switch_times_ISs_Case_GSA}
\begin{tabular}{cccc}
Switch Times & standard GSA & Shapley GSA \\
\hline
$T_1$ & 0.6579 & 0.2994\\
$T_2$ & 1.1374 & 0.3998\\
$T_3$ & 1.4381 & 0.9713\\
$T_4$ & 1.7447 & 1.5756\\
$T_5$ & 2.1806 & 2.3869\\
$T_6$ & 2.6691 & 2.7090\\
$T_7$ & 3.1736 & 3.2766\\
$T_8$ & 3.7898 & 3.7898\\
\hline
\hline
Magnitudes & standard GSA & Shapley GSA  \\
\hline
$A_0$ & 0.2175 & 0.1215\\
$A_1$ & 0.1519 & 0.1067\\
$A_2$ & 0.0920 & 0.0433\\
$A_3$ & 0.0812 & 0.1839\\
$A_4$ & 0.0879 & 0.1437\\
$A_5$ & 0.1207 & 0.1543\\
$A_6$ & 0.0956 & 0.0546\\
$A_7$ & 0.0914 & 0.0896\\
$A_8$ & 0.0618 & 0.1024\\
\hline
\end{tabular}
\end{center}
\end{table}
Fig.~\ref{fig:06_GSA_final} shows $R_\alpha$ when the standard and Shapley-based GSA IS are applied. The variation of $R_\alpha$ is more aggressive in the standard GSA IS than in the Shapley-based GSA IS. Hence it is interesting to look at performance metrics, specifically mean and standard deviation, for the GSA-based IS.
Table~\ref{table:mean_and_std_GSA_TDF} illustrates the mean and standard deviation of the GSA shapers. It can be seen that the mean for the standard GSA IS is already much lower than that of any controller previously shown in Table~\ref{table:mean_and_std_classic_shapers}. This shows the clear advantage of a GSA-optimized IS. Also, the standard deviation $\sigma$ is much smaller with $0.8959$ and $0.6097$ for $\alpha$ and $\beta$, respectively, than its counterparts for any other controller in Table~\ref{table:mean_and_std_classic_shapers}. The Shapley value for $Sh^{R_\alpha}_L$ of the standard GSA IS is $0.8850$. 

We enforce with the Shapley-based GSA IS to constrain the influence of $L$ toward the range of $R_\alpha$ to be $\leq 0.85$. By reducing the sensitivity of $R_\alpha$ to the cable length uncertainty, the variance of the swing angle also shrinks. Table~\ref{table:mean_and_std_GSA_TDF} shows for $R_\alpha$ that the mean is increasing compared to the standard GSA IS but the deviation is shrinking to $0.7807$, because the influence of $L$ is minimized. This shows that by enforcing the Shapley value constraint, the standard deviation with respect to an uncertain parameter can be minimized. As such, Shapley-value-based constraints can provide a means to mitigate sensitivity of certain performance objectives to such uncertainty. Note that this desensitization comes at a cost to the $\mu$ and $\sigma$ for the swing angle $\beta$, which is expected for a nonlinear, nonconvex optimization with multiple objectives. 
\begin{table}
\begin{center}
\caption{Mean and standard deviation of the standard and Shapley-based GSA IS.}\label{table:mean_and_std_GSA_TDF}
\begin{tabular}{cccc}
 & Standard GSA & Shapley GSA \\
\hline
$\mu(R_\alpha)$ [deg] & 7.8303 & 7.9743\\
$\sigma(R_\alpha)$ [deg] & 0.8959 & 0.7807\\
$\mu(R_\beta)$ [deg] & 4.7433 & 7.6740 \\
$\sigma(R_\beta)$ [deg] & 0.6097 & 2.2978\\
\hline
\end{tabular}
\end{center}
\end{table}
Table~\ref{table:Shapley_values_GSA_ISs} presents the Shapley values of the swinging angle ranges $R_\alpha$ and $R_\beta$ for both the standard and the Shapley-based GSA IS. The results show that the constraint $Sh^{R_\alpha}_L \leq 0.85$ is satisfied, and that $Sh^{R_\alpha}_{m_l}$ is increased compared to the standard GSA IS. It should be noted that, analytically, Shapley values are expected to sum to 1. However, since this study is conducted numerically on a nonlinear dynamical system, exact summation cannot be guaranteed, nor can it be ensured that all Shapley values are non-negative without increasing computational resources devoted to simulation. Based on the authors’ experience in ~\cite{stein_shapley_2025}, these deviations can be mitigated by increasing the number of Monte Carlo simulations or by developing improved surrogate models for the UAV-payload system, accounting for uncertainties in rope length and payload mass. Despite these limitations, the authors assert that the presented results provide a promising new approach for precision motion control in aerial payload transport.
\begin{table}
\begin{center}
\caption{Shapley Values for the GSA ISs.}\label{table:Shapley_values_GSA_ISs}
\begin{tabular}{ccccc}
Metric & Standard GSA & Shapley GSA \\
\hline
$Sh^{R_\alpha}_{m_l}$ & 0.1090 & 0.1969\\[3pt]
$Sh^{R_\alpha}_L$ & 0.8850 & 0.8500\\[3pt]
$Sh^{R_\beta}_{m_l}$ & 0.0256 & -0.0178\\[3pt]
$Sh^{R_\beta}_L$ & 0.9932 & 0.9640\\[3pt]
\hline
\end{tabular}
\end{center}
\end{table}
%
%
%
\section{OUTLOOK: EXPERIMENTS}
\label{sec:Experiment}

\begin{figure}[h]
\centering
	\includegraphics[width=0.4\textwidth]{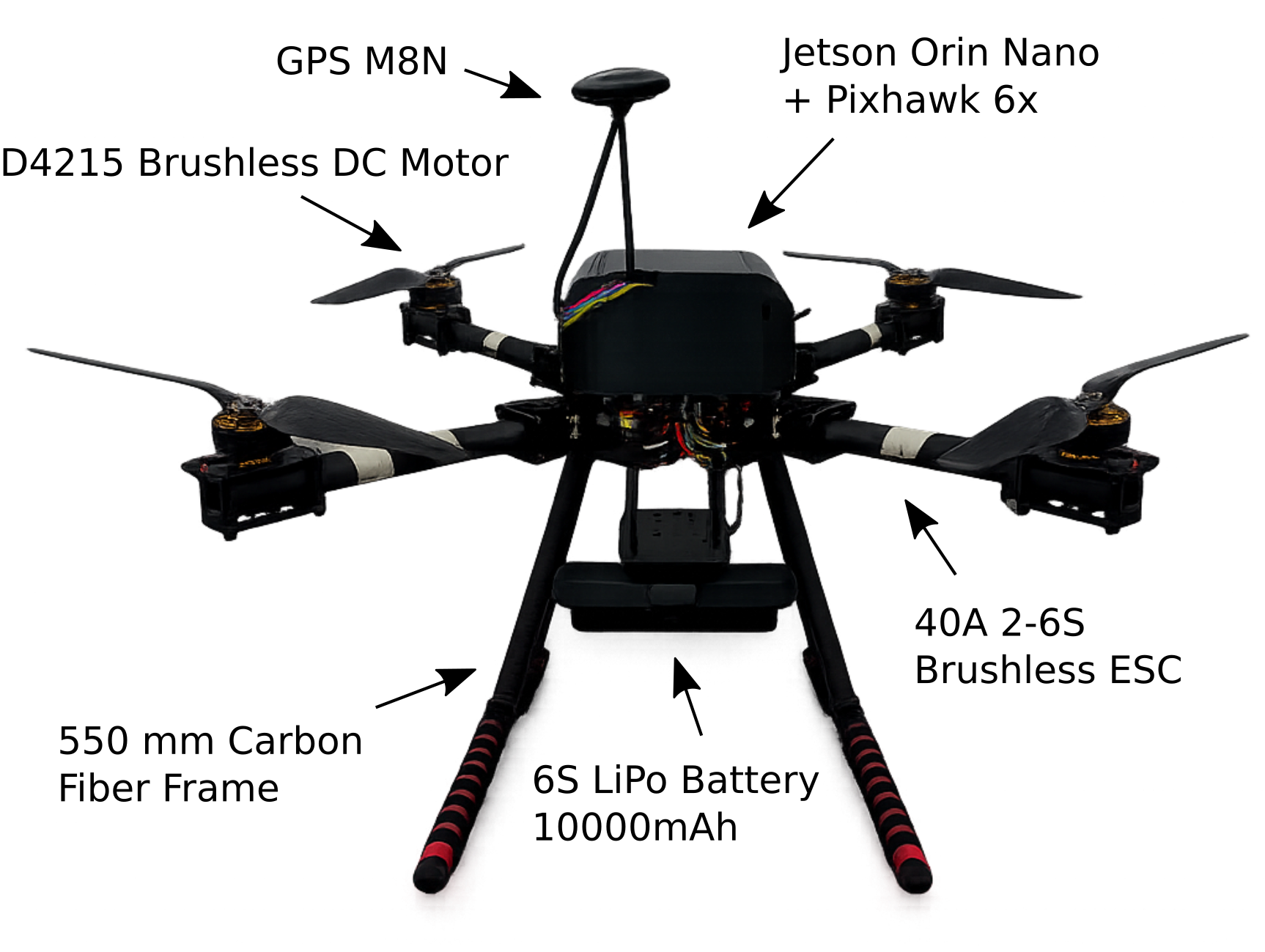}
	\caption{Drone for experiments.}
 \label{fig:7_Experimental_Drone}
\end{figure}

\begin{figure}[h]
\centering
	\includegraphics[width=0.4\textwidth]{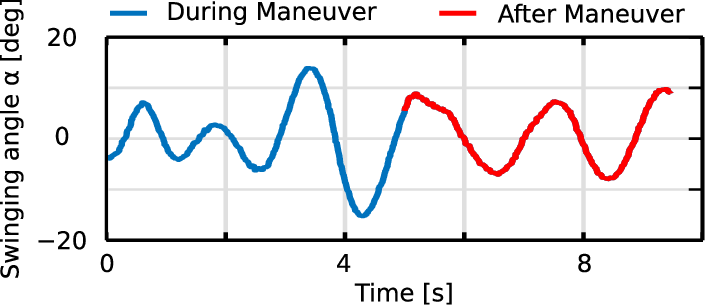}
	\caption{Experimental results for a pulse input.}
 \label{fig:8_Pulse_Signal_Experiments}
\end{figure}

To justify the usage of input shaping for this particular application, and specifically the TDF designed through Shapley-based global sensitivity analysis, the drone shown in Fig.~\ref{fig:7_Experimental_Drone} will be mounted with a payload of 0.6 kg, attached by a cable with a length of 1 m. For preliminary considerations to experimental data collection, a pulse profile for the system, with the drone maintaining a constant velocity of 5 m/s over 5 s in one direction, was collected, shown in Fig.~\ref{fig:8_Pulse_Signal_Experiments}. A ENC41270 rotary encoder with a resolution of 1000CPR was used to collect the swing angle of the payload. 

Future experimental work is reserved for applying the robust and GSA input shapers to this same system, and the pulse profile will serve as the benchmark for which the input shapers are judged against, as the shapers are expected to reduce the swinging angles. Furthermore, tests will also consider different masses for the payload, with 0.5 and 0.7 kg as the upper and lower bound. A cable length of 0.5 m will also be utilized for another portion of the tests, as the cable length as shown in Fig.~\ref{fig:06_GSA_final} has more effect on the swinging angles than the payload mass. This is to show a range of swing angles as the mass and cable are varied between tests.

%
%
%
\section{CONCLUSIONS}
\label{sec:conclusions}
This work demonstrates that GSA-based controller design can effectively desensitize UAV-payload motion controllers to system uncertainties without increasing maneuver time, providing a robust alternative to conventional methods. Incorporating Shapley values as constraints offered critical insights into handling parameter variability in the rope length and payload mass, though Monte Carlo simulations remain computationally intensive and using fewer samples results in numerical deviations. Future work will investigate the use of polynomial chaos expansion as a surrogate model to reduce computational cost, likely via a non-intrusive approach due to the system’s nonlinearity. Furthermore, as observed from desensitization of an uncertain parameter on a single objective, there is a tradeoff in the performance of other objectives. Studying the effects of different Shapley-based constraints for achievement of different application requirements will thus be vital to augmenting this approach. Most importantly, the authors plan to demonstrate experiments with the different input shapers to illustrate the feasibility of this work in hardware, as briefly introduced in Section ~\ref{sec:Experiment}. The developed desensitized precision motion controllers have broad applications, including high-accuracy aerial delivery, stable cinematography, precision surveying and mapping, infrastructure inspection, and payload stabilization under wind or turbulence. Experimental validation is planned to confirm the practical effectiveness of these strategies through different applications, specifically camera-oriented objectives and minimizing residual payload oscillations.

%
%
%
\bibliographystyle{IEEEtran}
\bibliography{references}

@article{stein_shapley_2025,
	title = {Shapley {Effects} as a {Global} {Sensitivity} {Metric} for {Robust} {Design} and {Control}},
	issn = {0143-2087, 1099-1514},
	doi = {10.1002/oca.3323},
	language = {en},
	urldate = {2025-09-09},
	journal = {Optimal Control Applications and Methods},
	author = {Stein, Adrian and Singh, Tarunraj},
	month = jun,
	year = {2025},
	pages = {oca.3323},
}

@article{ramli_pid_2025,
	title = {{PID} {Control} and {Input} {Shaping} for {Quadrotor} {UAV} {Stabilization} and {Payload} {Swing} {Reduction}},
	volume = {9},
	copyright = {Copyright (c) 2025 Liyana Ramli, Noor Hanis Izzuddin Mat Lazim, Amalin Aisya Mohd Awi, Aina Syazwin Mohd Shukor},
	issn = {2600-8084},
	language = {en},
	urldate = {2025-09-26},
	journal = {Applications of Modelling and Simulation},
	author = {Ramli, Liyana and Lazim, Noor Hanis Izzuddin Mat and Awi, Amalin Aisya Mohd and Shukor, Aina Syazwin Mohd},
	month = jun,
	year = {2025},
	pages = {264--272},
}

@article{akhtar_path_2022,
	title = {Path {Invariant} {Controllers} for a {Quadrotor} {With} a {Cable}-{Suspended} {Payload} {Using} a {Global} {Parameterization}},
	volume = {30},
	issn = {1558-0865},
	doi = {10.1109/TCST.2021.3133693},
	number = {5},
	urldate = {2025-09-26},
	journal = {IEEE Transactions on Control Systems Technology},
	author = {Akhtar, Adeel and Saleem, Sajid and Shan, Jinjun},
	month = sep,
	year = {2022},
	pages = {2002--2017},
}

@article{mohammadi_passivity-based_2022,
	title = {Passivity-{Based} {Control} of {Multiple} {Quadrotors} {Carrying} a {Cable}-{Suspended} {Payload}},
	volume = {27},
	issn = {1941-014X},
	doi = {10.1109/TMECH.2021.3102522},
	number = {4},
	urldate = {2025-09-26},
	journal = {IEEE/ASME Transactions on Mechatronics},
	author = {Mohammadi, Keyvan and Sirouspour, Shahin and Grivani, Ali},
	month = aug,
	year = {2022},
	pages = {2390--2400},
}

@article{zhu_modeling_2025,
	title = {Modeling, {Robust} {Control} {Design}, and {Experimental} {Verification} for {Quadrotor} {Carrying} {Cable}-{Suspended} {Payload}},
	volume = {22},
	issn = {1558-3783},
	doi = {10.1109/TASE.2024.3437747},
	urldate = {2025-09-26},
	journal = {IEEE Transactions on Automation Science and Engineering},
	author = {Zhu, Yang and Zheng, Zhiyuan and Shao, Jinliang and Huang, Hailong and Xing Zheng, Wei},
	year = {2025},
	pages = {6061--6075},
}

@article{stein_minimum_2023,
	title = {Minimum time control of a gantry crane system with rate constraints},
	volume = {190},
	urldate = {2025-09-26},
	journal = {Mechanical Systems and Signal Processing},
	publisher = {Elsevier},
	author = {Stein, Adrian and Singh, Tarunraj},
	year = {2023},
	pages = {110120},
}

@inproceedings{wang_dynamics_2016,
	title = {Dynamics modelling and linear control of quadcopter},
	issn = {2325-0690},
	doi = {10.1109/ICAMechS.2016.7813499},
	urldate = {2025-09-24},
	booktitle = {2016 {International} {Conference} on {Advanced} {Mechatronic} {Systems} ({ICAMechS})},
	author = {Wang, Pengcheng and Man, Zhihong and Cao, Zhenwei and Zheng, Jinchuan and Zhao, Yong},
	month = nov,
	year = {2016},
	pages = {498--503},
}

@inproceedings{james_cooperative_2024,
	address = {New York, NY, USA},
	series = {{DroNet} '24},
	title = {Cooperative {Drone} {Delivery} via {Push}-based {Lift} with {Payload} {Stabilization}},
	isbn = {979-8-4007-0656-1},
	doi = {10.1145/3661810.3663468},
	urldate = {2025-09-26},
	booktitle = {Proceedings of the 10th {Workshop} on {Micro} {Aerial} {Vehicle} {Networks}, {Systems}, and {Applications}},
	publisher = {Association for Computing Machinery},
	author = {James, Alice and Seth, Avishkar and Kuantama, Endrowednes and Han, Richard and Mukhopadhyay, Subhas},
	month = jun,
	year = {2024},
	pages = {31--36},
}

@article{han_controller_2022,
	title = {Controller {Design} and {Disturbance} {Rejection} of {Multi}-{Quadcopters} for {Cable} {Suspended} {Payload} {Transportation} {Using} {Virtual} {Structure}},
	volume = {10},
	issn = {2169-3536},
	doi = {10.1109/ACCESS.2022.3222031},
	urldate = {2025-09-26},
	journal = {IEEE Access},
	author = {Han, Xiao and Miyazaki, Ryo and Gao, Tianhua and Tomita, Kohji and Kamimura, Akiya},
	year = {2022},
	pages = {122197--122210},
}

@article{li_autotrans_2023,
	title = {{AutoTrans}: {A} {Complete} {Planning} and {Control} {Framework} for {Autonomous} {UAV} {Payload} {Transportation}},
	volume = {8},
	issn = {2377-3766},
	shorttitle = {{AutoTrans}},
	doi = {10.1109/LRA.2023.3313010},
	number = {10},
	urldate = {2025-09-26},
	journal = {IEEE Robotics and Automation Letters},
	author = {Li, Haojia and Wang, Haokun and Feng, Chen and Gao, Fei and Zhou, Boyu and Shen, Shaojie},
	month = oct,
	year = {2023},
	pages = {6859--6866},
}

@inproceedings{stein_aruco_2024,
	title = {{ArUco} {Based} {Reference} {Shaping} for {Real}-time {Precision} {Motion} {Control} for {Suspended} {Payloads}},
	urldate = {2025-09-26},
	booktitle = {2024 {American} {Control} {Conference} ({ACC})},
	publisher = {IEEE},
	author = {Stein, Adrian and Vexler, David and Singh, Tarunraj},
	year = {2024},
	pages = {4390--4395},
}

@article{shelare_payload_2024,
	title = {A payload based detail study on design and simulation of hexacopter drone},
	volume = {18},
	issn = {1955-2505},
	doi = {10.1007/s12008-023-01269-w},
	language = {en},
	number = {5},
	urldate = {2025-09-26},
	journal = {International Journal on Interactive Design and Manufacturing (IJIDeM)},
	author = {Shelare, Sagar and Belkhode, Pramod and Nikam, Keval Chandrakant and Yelamasetti, Balram and Gajbhiye, Trupti},
	month = jul,
	year = {2024},
	pages = {2675--2692},
}

@article{yazdannik_novel_2024,
	title = {A novel quadrotor carrying payload concept via {PID} with {Feedforward} terms},
	volume = {12},
	issn = {2049-6427},
	doi = {10.1108/IJIUS-10-2023-0141},
	number = {3},
	urldate = {2025-09-26},
	journal = {International Journal of Intelligent Unmanned Systems},
	author = {Yazdannik, Saman and Sanisales, Shamim and Tayefi, Morteza},
	month = may,
	year = {2024},
	pages = {331--347},
}

@incollection{kuhn_17_1953,
	title = {17. {A} {Value} for n-{Person} {Games}},
	isbn = {978-1-4008-8197-0},
	doi = {10.1515/9781400881970-018},
	urldate = {2025-09-26},
	booktitle = {Contributions to the {Theory} of {Games} ({AM}-28), {Volume} {II}},
	publisher = {Princeton University Press},
	author = {Shapley, L. S.},
	editor = {Kuhn, Harold William and Tucker, Albert William},
	month = dec,
	year = {1953},
	pages = {307--318},
}

@article{iooss_shapley_2019,
	title = {Shapley effects for sensitivity analysis with correlated inputs: comparisons with {Sobol}' indices, numerical estimation and applications},
	volume = {9},
	issn = {2152-5080, 2152-5099},
	doi = {10.1615/Int.J.UncertaintyQuantification.2019028372},
	language = {English},
	number = {5},
	urldate = {2025-06-13},
	journal = {International Journal for Uncertainty Quantification},
	publisher = {Begel House Inc.},
	author = {Iooss, Bertrand and Prieur, Clementine},
	year = {2019},
}

@article{muthusamy_real-time_2022,
	title = {Real-{Time} {Adaptive} {Intelligent} {Control} {System} for {Quadcopter} {Unmanned} {Aerial} {Vehicles} {With} {Payload} {Uncertainties}},
	volume = {69},
	issn = {1557-9948},
	doi = {10.1109/TIE.2021.3055170},
	number = {2},
	urldate = {2025-06-13},
	journal = {IEEE Transactions on Industrial Electronics},
	author = {Muthusamy, Praveen Kumar and Garratt, Matthew and Pota, Hemanshu and Muthusamy, Rajkumar},
	month = feb,
	year = {2022},
	pages = {1641--1653},
}

@article{eltayeb_integral_2022,
	title = {Integral {Adaptive} {Sliding} {Mode} {Control} for {Quadcopter} {UAV} {Under} {Variable} {Payload} and {Disturbance}},
	volume = {10},
	issn = {2169-3536},
	doi = {10.1109/ACCESS.2022.3203058},
	urldate = {2025-06-13},
	journal = {IEEE Access},
	author = {Eltayeb, Ahmed and Rahmat, Mohd Fua’ad and Basri, Mohd Ariffanan Mohd and Eltoum, M. A. Mohammed and Mahmoud, Magdi Sadek},
	year = {2022},
	pages = {94754--94764},
}

@article{yang_energy-based_2020,
	title = {Energy-{Based} {Nonlinear} {Adaptive} {Control} {Design} for the {Quadrotor} {UAV} {System} {With} a {Suspended} {Payload}},
	volume = {67},
	issn = {1557-9948},
	doi = {10.1109/TIE.2019.2902834},
	number = {3},
	urldate = {2025-06-13},
	journal = {IEEE Transactions on Industrial Electronics},
	author = {Yang, Sen and Xian, Bin},
	month = mar,
	year = {2020},
	pages = {2054--2064},
}

@article{geronel_adaptive_2025,
	title = {Adaptive sliding mode control for vibration reduction on {UAV} carrying a payload},
	volume = {31},
	issn = {1077-5463},
	doi = {10.1177/10775463241231845},
	language = {EN},
	number = {5-6},
	urldate = {2025-06-13},
	journal = {Journal of Vibration and Control},
	publisher = {SAGE Publications Ltd STM},
	author = {Geronel, Renan S and Bueno, Douglas D},
	month = mar,
	year = {2025},
	pages = {721--737},
}

@article{imran_adaptive_2024,
	title = {Adaptive {Control} of {Unmanned} {Aerial} {Vehicles} with {Varying} {Payload} and {Full} {Parametric} {Uncertainties}},
	volume = {13},
	copyright = {http://creativecommons.org/licenses/by/3.0/},
	issn = {2079-9292},
	doi = {10.3390/electronics13020347},
	language = {en},
	number = {2},
	urldate = {2025-06-13},
	journal = {Electronics},
	publisher = {Multidisciplinary Digital Publishing Institute},
	author = {Imran, Imil Hamda and Wood, Kieran and Montazeri, Allahyar},
	month = jan,
	year = {2024},
	note = {Number: 2},
	pages = {347},
}

@article{ogunbodede_load_2022,
	title = {Load {Vibration} {Mitigation} in {Unmanned} {Aerial} {Vehicles} {With} {Cable} {Suspended} {Load}},
	volume = {1},
	issn = {2690-702X},
	doi = {10.1115/1.4053576},
	number = {034502},
	urldate = {2025-06-13},
	journal = {Journal of Autonomous Vehicles and Systems},
	author = {Ogunbodede, Oladapo and Singh, Tarunraj},
	month = feb,
	year = {2022},
}

@article{qian_guidance_2020,
	title = {Guidance and {Control} {Law} {Design} for a {Slung} {Payload} in {Autonomous} {Landing}: {A} {Drone} {Delivery} {Case} {Study}},
	volume = {25},
	issn = {1941-014X},
	shorttitle = {Guidance and {Control} {Law} {Design} for a {Slung} {Payload} in {Autonomous} {Landing}},
	doi = {10.1109/TMECH.2020.2998718},
	number = {4},
	urldate = {2025-06-13},
	journal = {IEEE/ASME Transactions on Mechatronics},
	author = {Qian, Longhao and Graham, Silas and Liu, Hugh H.-T.},
	month = aug,
	year = {2020},
	pages = {1773--1782},
}

@article{stein_global_2022,
	series = {17th {IFAC} {Workshop} on {Time} {Delay} {Systems} {TDS} 2022},
	title = {Global {Sensitivity} {Analysis} based {Design} of {Input} {Shapers}},
	volume = {55},
	issn = {2405-8963},
	doi = {10.1016/j.ifacol.2022.11.335},
	number = {36},
	urldate = {2025-06-13},
	journal = {IFAC-PapersOnLine},
	author = {Stein, Adrian and Singh, Tarunraj},
	month = jan,
	year = {2022},
	pages = {67--72},
}

@article{mahesh_fire_2023,
	series = {International {Conference} on {Startup} ventures: {Technology} {Developments} and {Future} {Strategies}},
	title = {Fire fighter drone with robotic gripper},
	volume = {79},
	issn = {2214-7853},
	doi = {10.1016/j.matpr.2022.12.027},
	urldate = {2025-06-13},
	journal = {Materials Today: Proceedings},
	author = {{Mahesh} and Baruah, Raktim Lal and {Krishan} and {Preeti} and Dagar, Sansh Bir},
	month = jan,
	year = {2023},
	pages = {334--337},
}

@article{singh_closed-form_2007,
	title = {Closed-form minimax time-delay filters for underdamped systems},
	volume = {28},
	copyright = {Copyright © 2006 John Wiley \& Sons, Ltd.},
	issn = {1099-1514},
	doi = {10.1002/oca.790},
	language = {en},
	number = {3},
	urldate = {2025-06-13},
	journal = {Optimal Control Applications and Methods},
	author = {Singh, Tarunraj and Muenchhof, Marco},
	year = {2007},
	note = {\_eprint: https://onlinelibrary.wiley.com/doi/pdf/10.1002/oca.790},
	pages = {157--173},
}

@inproceedings{nettekoven_3d_2021,
	title = {A {3D} {Printing} {Hexacopter}: {Design} and {Demonstration}},
	issn = {2575-7296},
	shorttitle = {A {3D} {Printing} {Hexacopter}},
	doi = {10.1109/ICUAS51884.2021.9476759},
	urldate = {2025-06-13},
	booktitle = {2021 {International} {Conference} on {Unmanned} {Aircraft} {Systems} ({ICUAS})},
	author = {Nettekoven, Alexander and Topcu, Ufuk},
	month = jun,
	year = {2021},
	pages = {1472--1477},
}

@inproceedings{hunt_3d_2014,
	title = {{3D} printing with flying robots},
	issn = {1050-4729},
	doi = {10.1109/ICRA.2014.6907515},
	urldate = {2025-06-13},
	booktitle = {2014 {IEEE} {International} {Conference} on {Robotics} and {Automation} ({ICRA})},
	author = {Hunt, Graham and Mitzalis, Faidon and Alhinai, Talib and Hooper, Paul A and Kovac, Mirko},
	month = may,
	year = {2014},
	pages = {4493--4499},
}

@book{singh_optimal_2009,
	address = {Boca Raton},
	title = {Optimal {Reference} {Shaping} for {Dynamical} {Systems}: {Theory} and {Applications}},
	isbn = {978-0-429-07541-4},
	shorttitle = {Optimal {Reference} {Shaping} for {Dynamical} {Systems}},
	doi = {10.1201/9781439805633},
	publisher = {CRC Press},
	author = {Singh, Tarunraj},
	month = oct,
	year = {2009},
}

\end{document}